\documentstyle[12pt]{article}

\newcommand{\be}{\begin{equation}}
\newcommand{\ee}{\end{equation}}
\newcommand{\ba}{\begin{array}{l}}
\newcommand{\ea}{\end{array}}
\newcommand{\bea}{\begin{eqnarray}}
\newcommand{\eea}{\end{eqnarray}}
\newcommand{\bb}{\begin{thebibliography}}
\newcommand{\eb}{\end{thebibliography}}

\newcommand{\nn}{\nonumber}

\begin{document}
\begin{titlepage}
\begin{center}
{\bf   RESEARCH INSTITUTE OF APPLIED PHYSICS \\
       TASHKENT STATE UNIVERSITY} \\
\vspace{20mm}
\hfill{\bf Preprint NIIPF-93/07}

\vspace{60mm}
\begin{center}{\bf QUANTUM DEFORMATIONS FOR THE DIAGONAL R-MATRICES\\
Talk at Workshop on Supersymmetry and Quantum Groups, Dubna, 15-20 July 1993}
\end{center}

\vspace{20mm}
 {\bf B.M.ZUPNIK} \\
\vspace{70mm}
{\bf Tashkent - 1993}
\end{center}
\end{titlepage}

\begin{center}{\bf QUANTUM DEFORMATIONS FOR THE DIAGONAL R-MATRICES
}\end{center}
\vskip 10mm
\begin{center}{\large{B.M.ZUPNIK}}\end{center}
\begin{center}{Research Institute of Applied Physics, Tashkent State
University, Vuzgorodok, Tashkent 700095, Uzbekistan}\end{center}
 \vskip 15mm
\begin{abstract}
We consider two different types of deformations for the linear group $ GL(n)$
which correspond to using of a general diagonal R-matrix. Relations between
braided and quantum deformed algebras and their coactions on a quantum plane
are discussed. We show that tensor-grading-preserving differential calculi
can be constructed on braided groups , quantum groups and quantum planes for
the
case of the diagonal R-matrix.

\end{abstract}
\vskip 20mm

     Deformations of the linear group $ GL(n)$, vector spaces and commutative
algebras are intensively discussed in the theory of quantum groups [1-4].
More simple $(G,f)$-graded deformations of Lie algebras are well known [5,6]
and recently their R-matrix generalizations (braided algebras) have been
considered [7,8].

We shall discuss the simplest multiparametric deformations of the commutative
algebra and linear group generating by a diagonal $R_D$-matrix [9]. The
general diagonal unitary R-matrix has the following form:
\be
(R_D )_{km}^{ij} =q(ji)\delta^i_k \delta^j_m
\label{a1}
\ee
where $q(ji)$ are complex deformation parameters
\be
q(ji)=1/q(ij) \; ,\;\;\; q(ii)=1
\label{a2}
\ee

This R-matrix is a partial case $X=1$ of the multiparametric nondiagonal
R-matrix [10].

Let us define a quantum plane [2] as the formal-series algebra with generators
$x^i \;(i=1,2\ldots n)$
\be
x^i\; x^j =(R_D )_{km}^{ji}\;x^k\;x^m = q(ij)x^j\;x^i
\label{a3}
\ee

One can treat the deformed groups as different cotransformations on the
quantum plane. We shall study two types of $GL(n)$ deformations using
$R_D$-matrix (1): braided group $DGL(n)$ and quantum group $GL_D (n)$.
Symbol $D$ in our notations corresponds to the diagonal structure of
$R_D$-matrix.

One can use the natural tensor $I_D$-grading on the quantum plane and
corresponding differential calculus. Here $I_D$ is a commutative semigroup,
which can be defined as the set of covariant and contravariant multiindices
with the multiplication rule $\ast $
\be
I\left(\frac{p}{r}\right)=\left(\frac{i_1\ldots i_p}{k_1\ldots k_r}\right)\;,
\;\;\;I\left(\frac{p}{r}\right)\ast I\left(\frac{s}{t}\right)=I\left(\frac{p+s}
{r+t}\right)
\label{a4}
\ee

The order of upper or low indices is unessential for the tensor grading so we
can use some type of symmetrization of indices, for instance,
D-symmetrization.
Note, that we must distinguish the multiindices with equal number indices
of different sort , e.g. $  (123)\neq (345)$. Nevertheless one can consider
the equivalence relation $E$ in the set $I_D$:{\bf the multiindices are
equivalent if they have some identical subset of indices and also an
arbitrary
number   of upper and low indices  coinciding in pairs } , for example
\be
{\scriptstyle \left(\frac{12}{34}\right)\sim \left(\frac{121}{341}\right)
\sim \left(\frac{1235}{3435}\right) \sim \left(\frac{12357}{34357}\right)}
\label{a5}
\ee

Let us denote $\widetilde{I}_D =I_D /E $ the corresponding factorset which
has also a multiplication rule for classes of multiindices .
Consider the generators
$ dx^i,\;\partial_i $ of the special differential calculus preserving
$\widetilde{I}_D $-grading [9]
\bea
& dx^i\;x^k=[ik]\;x^k\;dx^i & \nn\\
&  dx^i\;dx^k=-[ik]\;dx^k\;dx^i & \nn\\
& \partial_i\;x^k=\delta^k_i + [ki]\; x^k\;\partial_i &\label{a6}\\
&  \partial_i\; \partial_k=[ik] \partial_k\; \partial_i \nn\\
&  \partial_k\;dx^i=[ik]dx^i\; \partial_k \nn
\eea
Here the notation $ [ik]=q(ik)$ is used.

Let us introduce the following notations for $\widetilde{I}_D $-graded
algebras:
$ M_D (n)=M_D (x)$ is the algebra with generators $x^i $, $\Lambda_D (n) $ is
the external algebra with generators $x^i , dx^i $ and $ D_1 (M_D ) $ is a
tangent vector space of first-order differential operators on $ M_D (n)$.
$ D_1 (M_D ) $ can be treated as the infinite-dimensional Lie D-algebra with
the generalized D-commutator of basic elements [9]
\be
D_k^{I(p)}=x^{I(p)}\;\partial_k=x^{i_1}\cdots x^{i_p}\partial_k
\label{a7}
\ee

Consider a simple braided D-deformations of the linear group $DGL(n)$ and
 corresponding Lie algebra $dgl(n)$ [5,6,9]. $DGL(n)$ is an unital algebra
 with noncommuting generators $L^i_k $
 \be
L^i_k L^j_m=[ij][jk][km][mi] L^j_m L^i_k={\scriptstyle\left[\frac{i\mid j}
{k\mid m}\right]} L^j_m\;L^i_k
 \label{a8}
 \ee
 where a special notation for the commutation factor is used. $M_D (n)$ and
 $DGL(n)$ are examples of D-commutative algebras. A  structure of Hopf
 D-algebra on  $DGL(n)$ is consistent with $\widetilde{I}_D $-grading
 \bea
 &  \Delta(L^i_k)=L^i_j \otimes_q L^j_k=[ij][jk][ki]L^j_k\otimes_q
 L^i_kj         & \nn\\
 & s(L^i_k)=(L^{-1})^i_k\;,\;\;\varepsilon(L^i_k)=\delta^i_k &\label{a9}
 \eea
 where a special D-symmetrical tensor product is used [9]. We can choose the
 exponential parametrization for the braided matrices
 \bea
 &  L^i_k=(\mbox{exp}\lambda)^i_k=\delta^i_k + \lambda^i_k +\ldots
         & \label{a10}\\
 & \lambda^i_k=\lambda^j_m (M^m_j)^i_k \;,\;\;\;(M^m_j)^i_k=\delta^i_j
 \delta^m_k &
 \label{a11}
 \eea

 Here  $M^m_j$ are generators of the fundamental representation of $dgl(n)$
 and $ \lambda^j_m (M^m_j)$ is an element of the corresponding
 noncommutative envelope [9]. Note , that the matrix elements of
 $dgl(n)$-generators  are not deformed.

 Cotransformation of $DGL(n)$ on the quantum plane $M_D (n)$ conserves
 $\widetilde{I}_D $-grading
 \bea
  & x^{\prime i}= L^i_k\otimes_q x^k=[ik] x^k \otimes_q  L^i_k &
  \label{a12}\\
  & d x^{\prime i}= L^i_k \otimes_q dx^k &\label{a13}\\
  & \partial^{\prime}_i =\partial_k \otimes_q (L^{-1})^k_i &\label{a14}
  \eea

It is convenient to use the equivalence relations for the grading of matrices
  \be
  L^i_k \sim (L^2 )^i_k \sim (L^{-1})^i_k \sim (\mbox{ln}L)^i_k
  \label{a15}
  \ee

   One can impose an additional restriction on the deformation parameters
   $[ik]=q(ik)$ , which results in disappearance of deformations for
   $dgl(n)$
   \be
   [ik][kj]=[ij]\;\Longrightarrow\;{\scriptstyle\left[\frac{i\mid j}
   {k\mid m}\right]}=1
   \label{a16}
   \ee

 Note , that $dgl(2)\simeq gl(2)$ without any additional restriction. Using
 the Eq(16) one can define the coaction of commutative matrices $L^i_k$ on
 the noncommutative space $M_D (n)$
 \bea
 & L^i_k\;L^j_m =L^j_m\;L^i_k &
 \nn \\
 & L^i_k \otimes_q x^l =[il][lk] x^l  \otimes_q L^i_k&
 \label{a17}
\eea

Now we shall consider a quantum deformation $GL_D (n)$ of the linear group
for the diagonal R-matrix (1). $F(GL_D (n))$ is a Hopf algebra with
generators $ a^i_k $
\be
R_D A_1 A_2 =A_2 A_1 R_D \Longrightarrow\;a^i_k\;a^j_m=[ij][km]a^j_m\;a^i_k
\label{a18}
\ee

Stress , that multiplication and comultiplication in $F(GL_D (n))$ conserve
some tensor grading , but an antipode map change tensor grading
 \bea
 & s(a^i_k ) =(a^{-1})^i_k &
 \nn \\
 & (a^{-1})^i_k \;(a^{-1})^j_m =[ji][km](a^{-1})^j_m\;(a^{-1})^i_k &
 \label{a19}\\
 & a^i_k \;(a^{-1})^j_m =[mi][kj](a^{-1})^j_m\;a^i_k &
 \nn
\eea

The $\widetilde{I}_D $-differential calculus on $M_D (n)$ (6) is covariant
with respect to the coaction of the quantum group $GL_D (n)$
\bea
  x^{\prime i}= a^i_k \otimes x^k & , & dx^{\prime i}= a^i_k \otimes
  dx^k
 \nn \\
 &\partial^{\prime}_i =\partial_k \otimes (a^{-1})^k_i=
 (a^{-1})^k_i \otimes \partial_k
 &\label{a20}
 \eea
 where the symmetrical tensor product is used.

Stress , that different deformations of the linear group are closely
connected as different cotransformations of the quantum plane . Let us
discuss the manifestations of this connection and examples of the combined
using of $GL_D (n)$ and $DGL(n)$ structures :

A) Consider a differential calculus on the quantum group $GL_D (n)$
\bea
& da^i_k\;a^j_m = [ij][mk] a^j_m\;da^i_k&
\nn \\
& da^i_k \;da^j_m =-[ij][mk] da^j_m\;da^i_k &
\label{a21}\\
& da^i_k \;(a^{-1})^j_m =[mi][kj](a^{-1})^j_m\;da^i_k &
\nn
\eea

Right-invariant differential forms $\rho^i_k $ on $GL_D (n)$ [3] can be
treated as 1-forms with $dgl(n)$-structure
\bea
& \rho^i_k = da^i_l (a^{-1})^l_k & \nn \\
& \rho^i_k\;\rho^j_m = -\left[\frac{i\mid j}{k\mid m}\right] \rho^j_m\;
  \rho^i_k & \label{22}
  \eea

B) $DGL(n)$-matrices generate "left orbit" on the quantum $GL_D (n)$-matrices
\bea
  & a^i_k \rightarrow \widetilde{a}^i_k=L^i_j\;a^j_k &
  \nn \\
  & L^i_l \;a^r_s =[ir][rl] a^r_s\;L^i_l &
 \nn \\
&  \widetilde{a}^i_k\;\widetilde{a}^j_m = [ij][mk] \widetilde{a}^j_m\;
\widetilde{a}^i_k &\label{a23} \\
  & \widetilde{a}^i_k\;a^j_m =[ij][mk] a^j_m \;\widetilde{a}^i_k& \nn \\
& L^i_k =\widetilde{a}^i_j\;(a^{-1})^j_k \nn
\eea
Note , that $DGL(n)$-matrices can be constructed in terms of quantum matrices.

C) Quantum $GL_D$-matrices can define "two-sided quantum orbit" on $DGL(n)$
\be
L^i_k \leftarrow a^i_j\;L^j_m\;(a^{-1})^m_k
\label{a24}
\ee
where the elements $a^i_j , (a^{-1})^m_k $ commute with $L^i_m$. This
construction was used in Ref[11] for deformation of gauge fields .

Deformations of the special linear group $SL(n)$ for the diagonal
$R_D$-matrix can be obtained by a factorization of quantum $GL(n)$-matrices
[12,13]
\be
a^i_k=\hat{a}^i_k (D_q )^{1/n} \label{a25}
\ee
where $D_q $ is the quantum determinant and $\hat{a}^i_k$ are generators
of $SL_D (n)$-group . Note , that a quantum deformation vanishes for
$SL_D (2)$ group , however we have nontrivial one-parametric deformation
for the case of $SL_D (3)$ quantum group.

Let $ q({\scriptstyle 12}) , q({\scriptstyle 23}) , q({\scriptstyle 31})$
 are independent deformation parameters of
$GL_D (3)$ (18). Consider a quantum determinant of $GL_D (3)$-matrix $a^i_k$
\bea
& D_q =a^1_1 a^2_2 a^3_3 +{\scriptstyle [21][31]} a^1_2 a^2_3 a^3_1 +
{\scriptstyle [31][32]} a^1_3 a^2_1 a^3_2 - & \nn \\
&{\scriptstyle [21]} a^1_2 a^2_1 a^3_3 - {\scriptstyle [32]} a^1_1 a^2_3
a^3_2 -{\scriptstyle [21][31][32]} a^1_3 a^2_2 a^3_1 &
\label{a26}
\eea

The quantum determinant is not central element in  $GL_D (3)$. Eq(25)
determines elements of the $SL_D (n)$-matrix $\hat{a}^i_k $ which satisfy
the commutation relations (18) with the following deformation parameter
\be
\hat{q}({\scriptstyle 12})=\hat{q}({\scriptstyle 23})=\hat{q}
({\scriptstyle 31})=({\scriptstyle [12][23][31]})^{1/3}
\label{27}
\ee

\end{document}